\documentclass[prl,a4paper,superscriptaddress,twocolumn,amsmath,amssymb]{revtex4-1}

\usepackage{graphicx}
\graphicspath{{./Figs/}}
\usepackage{color}
\usepackage{bm}
\usepackage{dsfont}
\usepackage{epstopdf}
\usepackage{hyperref}
\usepackage{setspace}
\usepackage{soul}
\begin{document}

\title{Magnetic Tetrastability in a Spin Chain}

\author{Vivien Pianet}
\email{vivien.pianet@u-bordeaux.fr}
\affiliation{University of Bordeaux, CRPP, UPR 8641, F-33600 Pessac, France}
\affiliation{CNRS, CRPP, UPR 8641, F-33600 Pessac, France}
\affiliation{University of Bordeaux, IMB, UMR 5251, F-33400 Talence, France}
\affiliation{INRIA, F-33400 Talence, France}

\author{Matias Urdampilleta}
\affiliation{University of Bordeaux, CRPP, UPR 8641, F-33600 Pessac, France}
\affiliation{CNRS, CRPP, UPR 8641, F-33600 Pessac, France}

\author{Thierry Colin}
\affiliation{University of Bordeaux, IMB, UMR 5251, F-33400 Talence, France}
\affiliation{INRIA, F-33400 Talence, France}  
\affiliation{Bordeaux INP, IMB, UMR 5251, F-33400,Talence, France.}  

\author{Rodolphe Cl\'erac}
\affiliation{University of Bordeaux, CRPP, UPR 8641, F-33600 Pessac, France}
\affiliation{CNRS, CRPP, UPR 8641, F-33600 Pessac, France}

\author{Claude Coulon}
\email{coulon@crpp-bordeaux.cnrs.fr}
\affiliation{University of Bordeaux, CRPP, UPR 8641, F-33600 Pessac, France}
\affiliation{CNRS, CRPP, UPR 8641, F-33600 Pessac, France}

\date{\today}


\begin{abstract}
Bistability in magnetism has been extensively used, in particular to store information. Here we propose an alternative route by using tetrastable magnetic domains. Using numerical and analytical calculations we show that a spin chain with a canting angle of $\pi/4$ possesses four energy-equivalent states. We discuss the static properties of such chain such as the profile and the energy of the domain walls as they govern the relaxation of the magnetization. The realisation of such spin chain could enable the encoding of the information on four bits which is a potential alternative toward the increase of storage density. 
\end{abstract}

\maketitle

Spintronics using magnetic materials in electronic devices has made considerable progress from fundamental studies to practical applications\cite{Fert2008}. This technology is based on the discovery of magnetoresistive effects, such as the giant magnetoresistance in ferromagnetic conductors. Nowadays these properties are mainly used for reading information encoded in the magnetic domains of a hard drive disk\cite{Chappert2007}. Depending on the relative magnetization orientation of the domains (either up or down), a drastic change of the electrical resistance is observed in the read-head. The constant reduction of the domain size, which slightly approaches the domain wall thickness, has almost reached its limit in standard inorganic magnetic materials. Moreover, as the domain size reduces, the anisotropy and therefore the bistability of such system decreases \cite{Neel1955}. Hence, the challenge resides in finding new ways to store information on magnetic media. One of the approach consists in using alternative magnetic object such as molecular nanomagnets\cite{Caneschi1991,Sessoli1993} or single atoms\cite{Loth2012}, which are the smallest magnetic domains that one can create. 

Here, we propose another strategy which consists in going beyond the traditional bistable storage of magnetic information. We propose to use spin chain that exhibits magnetic tetrastability, that enables encoding of information on four states, and should largely extend the storage density. Using numerical and analytical calculations, we show that the so-called $\pi/4$ canted spin chain presents four stable magnetic domains with orthogonal magnetizations. Finally we show that domain walls, which are responsible for the relaxation of the magnetization (loss of information), have a finite energy which should avoid their nucleation at low temperature and therefore preserve the encoded information.
\begin{figure}[!h]%
\includegraphics[width=\columnwidth]{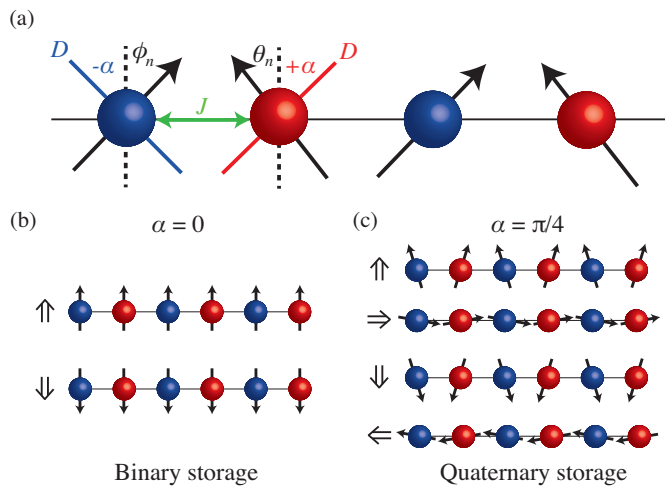}%
\caption{Information storage in a spin chain. (a) Representation of the spins orientations $\theta$ and $\phi$ associated to the easy axis $+\alpha$ and $-ˆ'\alpha$ (red and blue line). The angles $\theta_n$ and $\phi_n$ measure the orientation of the spins $\theta$ and $\phi$ relatively to the normal to the chain axis (dashed line). The $J$ and $D$ parameters describe respectively the magnetic interaction between two neighbouring spins and the magnetic anisotropy of a given spin. (b) Representation of the two energy-equivalent domains in the regular SCM, for  $\alpha=0$. (c) Representation of the four energy-equivalent domains for an $\alpha=\pi/4$ canted SCM for a finite $\theta_e$. 
}
\label{Fig1}%
\end{figure}

Among the variety of low dimensionnal magnets, single chain magnets (SCMs) \cite{Caneschi2001,Clerac2002}, have been extensively studied as they present a slow relaxation of magnetization, promising for information storage \cite{Caneschi2001,Clerac2002,Coulon2006,Coulon2015}. SCMs are generally made by assembling together single-molecule magnets that owns a strong uni-axial anisotropy\cite{Ferbinteanu2005}. A SCM can be simply described by a chain of spins, as depicted in Fig.\ref{Fig1}a, with the following parameters: $S$ is the amplitude of each spin, $D$ is the on-site magnetic anisotropy, $\alpha$ is the canting angle between the easy axis of magnetization and the normal to the chain axis, $J$ is the exchange interaction between two neighbouring spins  and $\theta_n$ and $\phi_n$ are the orientation of two consecutive spins at the site $n$ with canting angle $+\alpha$ and $-\alpha$.

In the simple case where the anisotropy axis of different sites are collinear ($\alpha=0$), the SCM presents a classical magnetic bi-stability: two kinds of magnetic domains exist with the same energy but opposite magnetizations (See Fig. \ref{Fig1}b). They are constituted of a given number of spins that are aligned along the unique easy axis of magnetization of the system\cite{Clerac2002}. Therefore, the domains are separated by domain walls (DWs) in which the magnetization is rotated by $\pi$, and are quoted as $\pi$ DWs  \cite{Barbara1973,Barbara1994}. It is important to note that the width of these DWs depends on the ration $D/J$.
The spin chains can be treated with the classical anisotropic Heisenberg Hamiltonian \cite{Coulon2006}:

\begin{equation}
H_{\alpha=0}=-2JS^2\sum_{-\infty}^{+\infty}\vec{u}_n.\vec{u}_{n+1}	-DS^2\sum_{-\infty}^{+\infty}u_{n,z}^2
\label{eq_hamilton}
\end{equation}
Where $\vec{u}_n$ is the unitary vector that gives the orientation of the $n^{th}$ spin of the chain and $z$ is the direction of the anisotropy axis ($\alpha=0$). Using this Hamiltonian, it has been shown that , for $D/J>4/3$, the DW becomes strictly narrow (the DW is located between two antiparallel spins). It corresponds to the so-called Ising limit \cite{Barbara1973}.

\begin{figure}[t]%
\includegraphics[width=\columnwidth]{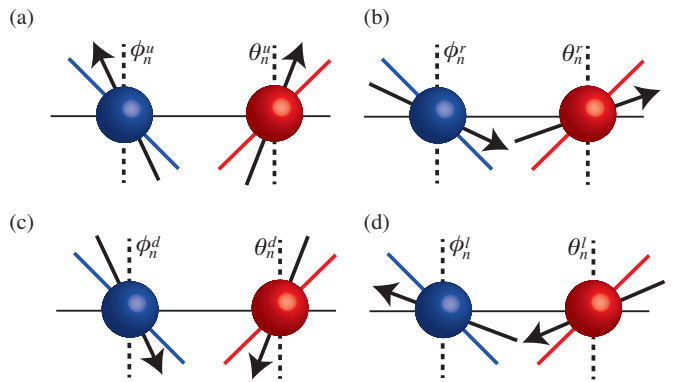}%
\caption{The four equilibirum configurations in the magnetic domains for a $D/J$ finite value. Configuration of the (a) "up", (b) "right", (c) "down" and (d) "left" magnetization.
}
\label{Fig2}%
\end{figure}

However, the magnetic topology of SCMs is generally more complex than the $\alpha=0$ regular case. Indeed, a large number of synthesized SCMs possesses two anisotropic axes with different orientations that alternate along the spin chain \cite{Barbara1971, Bernot2008, Bhowmick2012}. These systems are called canted SCM and own a finite angle $\alpha$. In these chains, the configuration of the spins in the magnetic domains is different from the $\alpha =  0$ case, due to a competition between the exchange interaction and the anisotropy in order to minimize the chain energy \cite{Coulon2015}.
 In the case of canted SCMs, the corresponding Hamiltonian is:
 \begin{equation}
\begin{split}
H_{\alpha\neq 0}&=-2JS^2\sum_{-\infty}^{+\infty}-cos(\phi_n-\theta_n)-cos(\theta_n-\phi_{n+1})\\
&+DS^2\sum_{-\infty}^{+\infty}sin^2(\phi_n+\alpha)-sin^2(\theta_n-\alpha)
\end{split}
\label{eq_H_chain}
\end{equation}
where $\theta_n$ and $\phi_n$ are the orientation of the spin associated to the easy axis $+\alpha$ and $-ˆ'\alpha$.
Using equation \ref{eq_H_chain} in the particular case $\alpha=\pi/4$, we can show that the energy of the chain is given by (See S.I.) :  
\begin{equation}
\begin{split}
\frac{E}{2JS^2}&=\sum_{-\infty}^{+\infty}-cos(\phi_n-\theta_n)-cos(\theta_n-\phi_{n+1})\\
&+\frac{D}{4J}\sum_{-\infty}^{+\infty}2+sin(2\phi_n)-sin(2\theta_n)
\end{split}
\label{eq_E_chain}
\end{equation}
In order to find the configuration of the system at the equilibrium we take the derivative of the chain energy with respect to $\theta_n$ and $\phi_n$ which leads to a system of angular equations: 
\begin{equation}
\begin{cases}
\frac{\partial E}{\partial\theta_n}=sin(\theta_n-\phi_n)+sin(\theta_n-\phi_{n+1})-\frac{D}{2J}cos(2\theta_n)\\
\\
\frac{\partial E}{\partial\phi_n}=sin(\phi_n-\theta_n)+sin(\phi_n-\theta_{n-1})+\frac{D}{2J}cos(2\phi_n)
\end{cases}
\label{eq_sys}
\end{equation}
At the equilibrium, the $\theta_n$ and $\phi_n$ angles are independent of the site number $n$ and are labelled $\theta_e$ and $\phi_e$. In these conditions, the summation of the equations of the system \ref{eq_sys} leads to the relation $cos(2\theta_e)=cos(2\phi_e)$. We find four solutions of lowest energy to this equation that lead to the four domains orientations described as following:

\begin{equation}
\begin{cases}
\theta^u_n=\theta_e,\phi^u_n=- \theta_e	
\\
\theta^r_n=\pi/2-\theta_e,\phi^r_n=\pi/2+ \theta_e		
\\
\theta^d_n=\pi+\theta_e,\phi^d_n=\pi- \theta_e	
\\
\theta^l_n=-\pi/2-\theta_e,\phi^l_n=-\pi/2+ \theta_e	
\end{cases}
\label{eq_solutions}
\end{equation}
Second, the equilibrium angle, $\theta_e$, can be deduced thanks to Eq. \ref{eq_sys}, and one of the solutions of Eq. \eqref{eq_solutions}:
\begin{equation}
tan(2\theta_e)=\frac{D}{4J}
\label{eq_E_thetaeq}
\end{equation}
These four configurations correspond to four domains with the same energy but different magnetizations, "up", "down", "right" and "left" as described in Fig.\ref{Fig2}. Therefore, the $\alpha=\pi/4$ canted spin chain can be viewed as a four-states system where each state corresponds to a domain with a specific magnetization orientation (Fig. 1c). 
In the following, we will describe $\pi/2$ DWs that separate two of these domains with orthogonal magnetization (e.g. "right" and "up"), as their profile and energy govern the static \cite{Nakamura1977, Nakamura1978} and dynamic \cite{Glauber1963, Billoni2011} properties of SCMs. For any value of the $D/J$ ratio, the profile of these $\pi/2$ DWs can be obtained numerically by solving Eq. \ref{eq_sys} using a Newton-Raphson refinement. The convergence of the refinement is reached when the energy of the DW (as defined in Eq. \ref{eq_E_DW}) is minimized (See S.I.).

\begin{equation}
\begin{split}
\frac{\Delta E}{2JS^2}&=\sum_{-\infty}^{+\infty}2cos(2\theta_e)-cos(\phi_n-\theta_n)-cos(\theta_n-\phi_{n+1})\\
&+\frac{D}{4J}\sum_{-\infty}^{+\infty}2 sin(2\theta_e)+sin(2\phi_n)-sin(2\theta_n)
\end{split}
\label{eq_E_DW}
\end{equation}

\begin{figure}[h!]%
\includegraphics[width=\columnwidth]{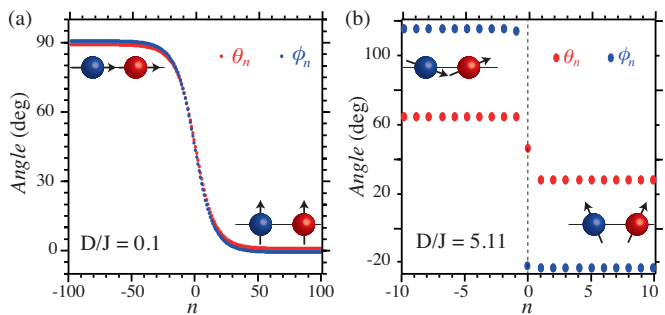}%
\caption{Profiles of $\pi/2$ domain walls in the broad ($D< J$, a) and sharp ($D> J$, b) limits for the $\alpha = \pi/4$ canted spin chain. The insets represent the right and up magnetization domains linked by the DW. The dashed line in the figure (b) shows that the $\pi/2$ DW center is pinned on a single spin.
}
\label{Fig3}%
\end{figure}

Fig. \ref{Fig3} (a) and (b) show typical $\pi/2$ DW profiles for $D<J$ and $D>J$ respectively.

In both cases, the DW links the "right" and "œup" magnetization domains. Its center is pinned on a $\theta$ site and corresponds to a spin perfectly aligned with its easy axis ($\theta_0=\pi/4$). As in the $\alpha=0$ case, the DWs are broad in the $D< J$ limit. However, for increasing $D/J$ ratio, the DW thickness decreases but stays still greater than the distance between two sites. As a consequence, and in contrast with the $\alpha=0$ case, strictly sharp DW (i.e. Ising-like) are forbidden in the $D\gg J$ limit.

In order to complete these numerical results, we derived analytically the DW profile in the two limits using Eq. (\ref{eq_sys}) and the DW energy (Eq. \ref{eq_E_DW}).
In the broad profile case ($D<J$), neighbouhring spins inside the DW have very close orientations. This fact leads us to introduce a continuous description of the DW profile, by defining the variables $\omega_n$ and $\gamma_n$:
\begin{equation}
\omega_n=\frac{\phi_n+\theta_n}{2}   \qquad   \gamma_n=\frac{\phi_n-\theta_n}{2}
\label{eq_variables_BDW}
\end{equation}
Thanks to these variables, the continuous calculation of equation (\ref{eq_E_DW}) can be carried out and the profile can be expressed as a function of the ratio $D/J$ (See SI):
\begin{equation}
\begin{cases}
\tan(\omega(u))=\exp(-u\frac{D}{J})
\\
\\
\gamma(u)=\frac{D}{8J}\frac{(1+\exp(-u\frac{D}{J}))}{\cosh(u\frac{D}{J})}-\theta_e
\end{cases}
\label{eq_profile_BDW}
\end{equation}
with $u$ the continuous variable describing the distance to the DW's center.
In the sharp profile case ($D>J$), the orientations of the spins inside the DW are very close to their equilibrium values due to the dominant magnetic anisotropy. Therefore, the system (\ref{eq_sys}) is linearized with respect to the angles $\delta\theta_n$ and $\delta\phi_n$:
\begin{equation}
\delta\theta_n=\theta_n-\theta_e   \qquad   \delta\phi_n=\phi_n+\theta_e
\label{eq_variables_SDW}
\end{equation}
Considering that, for a $\pi/2$ DW between "right" and "up" domains, the spin orientation $\theta_0$ is exactly equal to $\pi/4$, the profile can be calculated as an exponential decrease from the first spin after the DW's center (See SI):
\begin{equation}
\begin{cases}
\delta\theta_n = \delta\phi_1 \exp(-(n+\frac{1}{2})\psi)
\\
\delta\phi_n = \delta\phi_1 \exp(-(n-1)\psi) 
\end{cases}
\label{eq_profile_SDW}
\end{equation}
with $\psi$ the parameter define by $\cosh(\frac{\psi}{2})=\frac{D^2}{8J^2}+1$.
 
\begin{figure}[h!]%
\includegraphics[width=\columnwidth]{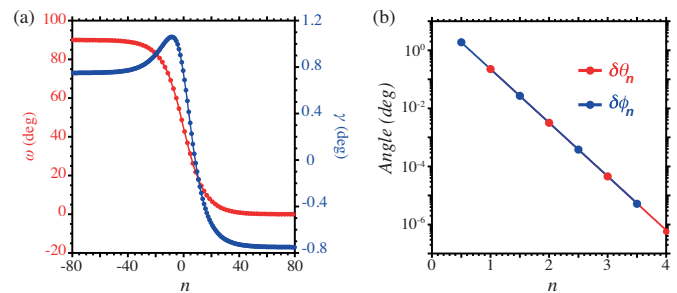}%
\caption{Comparison of the analytical expressions (\ref{eq_profile_BDW}) and (\ref{eq_profile_SDW}) with the numerical profile of the $\pi/2$ domain walls presented in Fig \ref{Fig3} in the broad ($D< J$, a) and sharp ($D> J$, b) limits. The dots are the numerical values of the profile variables and the lines are the fits of the same data with the analytical profile expressions. The analytical profile (a) leads to a $D/J$ value of 0.104735 in good agreement with the value obtained numerically of 0.104737. Similarly, the fit of profil (b) gives a $D/J$ value of 5.103 in good agreement with the numerical value of 5.111.
}
\label{Fig4}	
\end{figure}

The Fig. \ref{Fig4}(a) and (b) present the numerical calculation of DW profiles in the broad and sharp limits respectively, fitted with the analytical expressions  (\ref{eq_profile_BDW}) and (\ref{eq_profile_SDW}). 
For both broad and sharp limits the fitting of the variables $\omega_n$, $\gamma_n$ and $\delta\theta_n$, $\delta\phi_n$ respectively, are in good agreement with the values extracted from numerical computation. 

The agreement between the numerical profiles and their analytical expressions in the two limits validates our numerical method. As a consequence, this method is reliable to extrapolate the DW profiles between the two limits.

\begin{figure}[t]%
\includegraphics[width=\columnwidth]{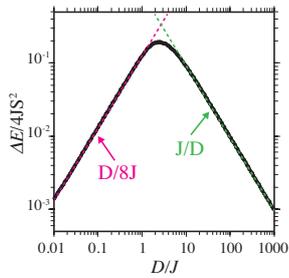}%
\caption{Energy of a $\pi/2$ domain wall as a function of the $D/J$ ratio. The dashed line represents the analytical limits of the normalized DW energy in the broad and sharp limits.
}
\label{Fig5}%
\end{figure}

After the description of the profile we have now to consider the energy associated to the DW. Indeed, if the DW has no energetic cost, their nucleations will relax the total magnetization of the SCM and therefore loose their information. 
Using the profile obtained from the Newton-Raphson refinement and equation \ref{eq_E_DW}, the energy of the $\pi/2$ DW is plotted in Fig. \ref{Fig5}. In both limits $D\ll J$ and $D\gg J$, the normalized energy follows simple power laws of the ratio $D/J$: 
\begin{equation}
\frac{\Delta E_{D<<J,\alpha=\pi/4}}{4JS^2}=\frac{D}{8J}\\
\label{eq_DE_broad}
\end{equation}
\begin{equation}
\frac{\Delta E_{D>>J, \alpha=\pi/4}}{4JS^2}=\frac{J}{D}
\label{eq_DE_sharp}
\end{equation}
Using the analytical profile expression obtained previously (Eq. \ref{eq_profile_BDW} and \ref{eq_profile_SDW}), these two limits can also be obtained analytically by solving Eq. \ref{eq_E_DW} in the continuous limit for $D\ll J$, and using its Landau development for $D\gg J$ (See SI). Between these two limits, the normalized DW energy exhibits a maximum of the order of $JS^2$. As a consequence, the DW has a finite energy which preclude their nucleation at sufficiently low temperature ($kT\ll 4JS^2$ for $D/J \sim 2.5$ ).

In conclusion, we have theoretically investigated the existence of four energy-equivalent states in the $\alpha=\pi/4$ canted spin chain. In the spin chain, these states are associated to four kind of magnetic domains which bear magnetization with different orientations. This striking property reinforces the potential interest of SCMs for data storage applications with the possibility to code the information on four states. In the same time, we emphasized the fact that these magnetic domains are linked by $\pi/2$ DWs instead of the $\pi$ DWs usually describe in the $\alpha\neq\pi/4$ case. The physical ingredients that describe the passage from $\pi$ to $\pi/2$ DWs when the canting angle get close to $\pi/4$ will be presented in a future report.  Moreover, we have determined the profile and the energy of the $\pi/2$ DWs thanks to numerical and analytical calculations. These results indicate that the highest DW energy is obtained when the $D/J$ ratio range between 1 and 10. This theoretical work opens numerous perspectives in the field of coordination chemistry and molecular magnetism in order to obtain experimental $\alpha=\pi/4$ canted SCMs.

This study has been carried out with financial support from the CNRS, the Universit\'e de Bordeaux, the region Aquitaine and the Agence National de la Recherche in the frame of the "Investments for the future" Programme IdEx Bordeaux - CPU (ANR-10-IDEX-03-02).

\bibliography{library}

\end{document}